\documentclass[twocolumn]{aastex631}

\usepackage{amsmath}
 \usepackage{graphicx}
\begin{document}

\title{Predictions for a Low-mass Cutoff for the Primordial Black Hole  Mass Spectrum}

\author{James Barbieri}
\affiliation{Advanced Systems Development,\\
1 Administrative Circle, \\
China Lake, 93555, USA - retired}

\author{George Chapline}
\affiliation{Space Science Institute, \\
Lawrence Livermore National Laboratory, \\
7000 East Ave., Livermore, CA}
%\email{chapline1@llnl.gov}
\date{}

\begin{abstract}
In this note we outline how a modest violation in the conservation of mass during the merger of two PBHs affects the PBH  mass spectrum that we previously obtained using a Boltzmann equation model for the evolution of the mass spectrum with no mass loss. We find that if the initial cosmological redshift is on the order 10$^{12}$, then the fraction of primordial holes with masses greater than $10^{3}$ solar masses appears  be close to what is required to provide the seeds for galaxies. In addition we note that as a result of rapid collisions and strong coupling to electromagnetic radiation for temperatures $>$ GeV \citep{Chapline2018}, there will be  an effective low mass cutoff in the mass spectrum for PBH masses less than a certain PBH mass less than than $0.1M_{\odot}$. We also point out that this cutoff in the mass spectrum below $\sim 0.1 M_\odot$ can be confirmed  by combining future microlensing observations from the Roman Space Telescope and the Vera C. Rubin Observatory with astrometric observations.
\end{abstract}

\keywords{gravitational collapse; cosmology; dark matter; vacuum energy; CMB.}

\section{Introduction} \label{sec:intro}

Primordial black holes (PBHs) provide an attractive explanation for dark matter in that they can not only provide the seeds for both galaxies and the large scale density inhomogenieties in the observed universe, but because BHs can possess a large entropy, then PBHs may explain the origin of the CMB \citep[see e.g.,][for a review]{Green2021}. In particular, if the cosmological redshift extends to $z\sim 10^{12}$ and the  initial mass spectrum for PBHs is restricted to masses between $0.01 M_\odot$ and $<1 M_\odot$, then it appears the appearance of a CMB and large scale inhomogenous structure of the universe would necessarily follow \citep{Chapline2019}. The signature for this scenario is that today there is a sharp cutoff in the PBH mass spectrum somewhere in the range $0.01-0.1M_{\sun}$.
In a previous paper \citep{Chapline2018}, we introduced a Boltzmann equation formalism for evaluating how the mass spectrum of PBHs evolves with cosmological time as a result of collisions and coalescence between black holes during the long period  $z<10^{10}$ when the standard cosmological model containing radiation and dark matter is applicable. A notable consequence  of our Boltzmann equation model and the assumption that dark matter consists entirely of PBHs is that the fraction of the dark matter PBH which appears with masses $>1000M_\odot$ is completely consistent with the assumption that these large mass PBHs became the seeds for galaxies and stellar clusters. In this paper we consider how our previous model for the PBH mass spectrum is modified if during mergers the sum of the masses of the merging BHs is reduced due to gravitational radiation or radiation of entropy \citep{Chapline1975, Chapline2001}.\\ 

Collisions between black holes in which the black holes do not coalesce will also result in a change in their masses due to gravitational radiation and entropy loss. However these mass losses will be much greater when the black holes actually coalesce. In this paper we neglect this effect, which significantly simplifies the Boltzmann equation calculation. In particular, we only include the first term in Eq.1 which  describes how the fraction of primordial black holes (PBHs) with a mass between $M$ and $M+dM$ increases with time  and due to the coalescence of a black hole of mass $M'$  and $M - M' + \Delta M_{GR}$, where  $\Delta M_{GR}$ is the mass loss due to gravitational radiation plus entropy loss. \\

The existence of mass loss during mergers of BHs as a result of gravitational radiation or entropy loss are among the black hole transformations allowed by Christodoulou. \citep{Christodoulou1970}.  In addition the existence of mass loss during mergers of BHs appears to be consistent with LIGO observations \citep{Isi2019}. In addition a strong coupling between BHs and thermal radiation is expected for temperatures $>$GeV in a quantum theory of gravity \citep{Chapline2001}. In this paper the effect of mass loss due to either gravitational radiation or the creation and loss of BH entropy created during collisions is accounted for by simply introducing an ad-hoc mass loss during a merger. Since neither the amount of gravitational radiation or entropy generation associated with the merger of two black holes can be calculated due to a lack of detailed understanding of how  these mass losses depend on the initial conditions, we simply reran our Boltzmann equation model with the additional ad-hoc assumption that the final mass differs of the merging black holes by a deficit $\Delta M_{GR}$. We have considered values for $\Delta M_{GR}$ in the range of $15 \% - 30 \%$. since these mass changes seems to be a reasonable based on the LIGO data. In the future, observations of the  "ring-down" gravitational signatures in the LIGO data may provide a much better understanding of the mass loss magnitudes expected for mergers of BHs with masses $>M_{\odot}$. \\

\section{Boltzmann equation model for mass spectrum}

The Boltzmann equation suggested in  \citep{Chapline2018} is:

\begin{equation}
\begin{split}
\frac{dp(M)}{dt} &= 27\pi\nu(t)\frac{\rho_{DM}}{\bar{M}(t)}\bigg[ \\
                & \int^{M}_{M^*}\sigma_{cap}(M',M-M')p(M-M')p(M')dM' \\
                 &- p(M)\int^\infty_{M^*}\sigma_{cap}(M,'M)p(M')dM' \bigg]
\end{split}
\label{eq:boltzman}
\end{equation}

\noindent where $\rho_{DM}$ is the present day dark matter density, $\nu (t)$ is the virial velocity of the at time t and $M_{DM}$ is the initial dark energy star mass (at $z \approx 10^{12}$ ), and $\sigma_{cap}$ is the Bae cross section. Here time runs from a very early time corresponding to a redshift $1+z_r \approx 10^{11}$ to  $z = 0.0$. However, in practice we were not able to extend our numerical solution to Eq. 1 to $z<10^8$. The time dependence of the average relative velocity $\nu(t)$ is of course complicated. However as a rough approximation we will assume that $\nu^2(t)$ scales roughly as the as the inverse square root of the distance between PBHs. \\

In the Newtonian limit, the angular momentum $L=b \mu v_{\infty}$ where $b$ is the impact parameter, $\mu$ is the reduced mass, $v_{\infty}$ is the virial velocity $\nu/c$ and the energy $E$ is $E=(1/2)\mu v^{2}_{\infty}$. The critical impact parameter $b_{\rm crit}$ is

\begin{equation}
b_{crit} = \frac{L}{\mu v_\infty} = \frac{L}{\sqrt{2}\mu E},
\end{equation}

\noindent For unequal masses  $b_{crit}$ is \citep{Kovetz2017},

\begin{equation}
b_{crit} = \left ( \frac{340 \pi}{3} \frac{m_1m_2( m_1 + m_2)^5}{v^2_\infty} \right )^{1/7} \;\; ,
\end{equation}

\noindent and the capture cross section is $\sigma_{cap} = \pi b^2_{\rm crit}$.

Since the first term of the Boltzmann equation represents the increase in mass due to mergers,  it is the only term where mass loss due to gravitational radiation or entropy generation is significant. We define the mass loss term $M_t((j)$ as
\begin{equation}
M_t(j) = \sigma_{cap} (t) \rho(t) \;\;,
\end{equation}
\noindent where $\sigma_{cap}(t)$  is the Bae cross section at time $t$ and $\rho(t)$ is the density at time $t$. The change in the mass $\Delta M_t(j)$ is:

\begin{equation}
\Delta M_t(j)= \frac{pc_{loss}}{\delta t} M_t(j) \;\; ,
\end{equation}

\noindent where $M_t(j)$ is the mass array $M(j)$ at time t, and the change in mass at each time step $\delta t = t_i - t_{i-i}$. (In our implementation $\delta t = 0.0532 $ at $z = 1.0\times 10^{12}$ to $\delta t = 33.54$ at $z = 1.0 \times 10^9$), and $pc_{loss}$ is the estimated percent loss in mass. The updated mass due to mass loss is:

\begin{equation}
M_t (j)= M_{t-1}(j) -  \Delta M_{t-1} (j) \;\; ,
\end{equation}

\noindent our mass distributions were calculated using above mass loss scenario.  In the following we show the initial and final normalized probability density $P(M)$ for PBH mass distributions for the scenarios considered.  In the following calculation we assume that  the initial  mass was $M^* = 0.05 M_\odot$ or $M^* = 0.1 M_\odot$, both with an initial redshift $z_i = 1.0 \times 10^{12}$ our final redshift  $z_f =1.0 \times 10^{8}$ is limited by the increasing computational expense of extending the numerical solution farther in time because of the small merger rate at smaller redshifts. Indeed we have found that the spectrum changes very slowly after $z =10^{8}$. Thus it is possible to get an idea where our calculated mass spectrum is consistent with the observed abundance of seeds of galaxies as well as N-body simulations of the formation of matter homogenieties, even though our formal calculation doesn't extend to $z<1000$ where structures form. As a result of these constraints the mass range we have considered include masses  $0.01M_\odot < M < 1000.0M_\odot$. In this paper we assume a $15\%$ to $30\%$ for the first term only, since the change in the spectrum due to the second term is expected to be much smaller.

\section{Mass Loss and Entropy}

Hawking's area rule \citep{susskind2005} states that the total area of a group of black holes represents an entropy that never deceases. On the other hand, the mass of two black holes can decrease due to gravitational radiation, as well as the radiation of BH entropy as thermal electromagnetic radiation. However, the specific heat of black holes is very large \citep{Chapline2001} which implies that the temperature arising from the mergers of two BHs with masses greater than $1 M_\odot$ is very low. Therefore it is not expected that mass loss due to radiation of thermal entropy will be important for these masses. A similar conclusion follows from Hawking's area rule even though the specific heat implied by Hawking's entropy contradicts the requirement that the specific heat of any physical system with many degrees of freedom must be positive. \\

What is possible, though, is that at the cosmological red shifts we are contemplating (viz.$>10^{10}$). The contribution to mass loss of thermal radiation due to quantum effects can be important \citep{Chapline2001}. As a result we expect that large mass losses due to both gravitational radiation and thermal heat radiation will be very important for cosmological redshifts $>\sim10^{12}$, leading to a sharp cutoff in the mass spectrum for PBH masses somewhere below $1 M_\odot$.  For PBH mergers with mass greater than $M_\odot$, we do not expect  that  BH entropy loss  will lead to an observable mass loss, although  gravitational radiation will continue to be observable. \\

%The fact that mass increases are not observed in the LIGO events shows that the contribution of entropy generation to the mass of a black hole formed by the merger of two black holes \citep{Christodoulou1970} is not significant - at least for Stellar masses. Of course this is consistent with the fact that the kinetic energy of PBHs, say at $z < 1000$ will be very small compared with their mass-energy. In the case of redshifts very close to the Big Bang - say  $z>10^{10}$, the internal energy created by mergers  may be comparable or even larger than their rest-energy. In our calculation we will take this possibility into account by choosing an initial mass for the PBHs that is fixed by the ratio of the CMB energy density to the DM density extrapolated from the current time to the chosen initial initial cosmological redshift. After this initial redshift we neglect the effect of entropy generation and  mass loss during mergers.\\ 

\section{Results}

For the results plotted here we use as an estimate of the coefficient of in Eq. (\ref{eq:boltzman})

\begin{equation}
10^{12}(1 + z)^3 \nu(t) \frac{\rho_{DM}}{\bar{M}(t)} \left( \frac{G}{c^2} \right )^2 \; \; ,
\end{equation}

\noindent  In Fig 1 P(M) distributions for  $m^*=0.1$ and $m^*=0.05$ - assuming no mass loss - is shown. In Figures 2 and 3 the results of the Boltzmann mass loss calculation for $m^*=0.05M_\odot$ and $m^*=0.1M_\odot$ , for no loss,  $15 \%$ mass loss, and $30 \%$ mass loss .

\begin{figure}[h!]
\includegraphics[width=\columnwidth]{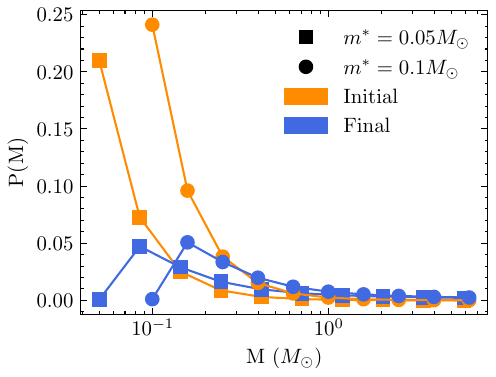}\caption{P(M) distributions - assuming no mass loss}
\end{figure}

\begin{figure}[h!]
\includegraphics[width=\columnwidth]{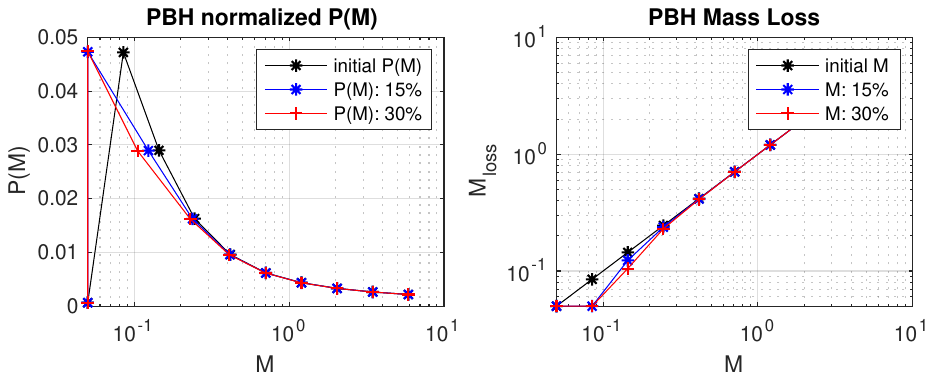}
\caption{P(M) for $m^*=0.05M_\odot$ for 15\% and 30\% mass losses}
\end{figure}

\begin{figure}[h!]
\includegraphics[width=\columnwidth]{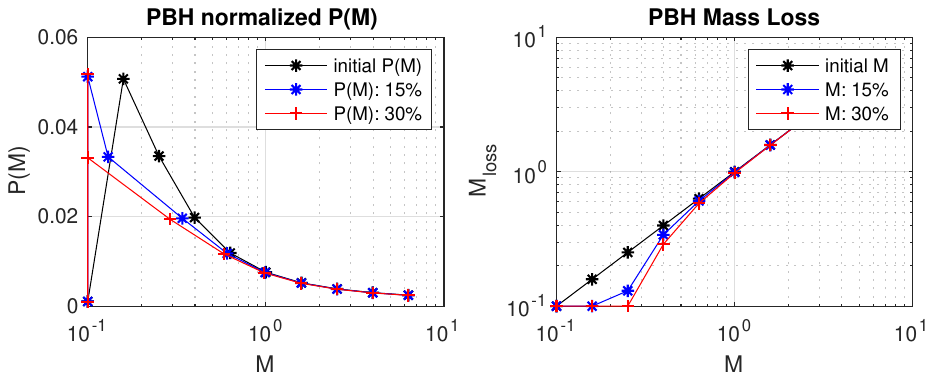}
\caption{P(M) for $m^*=0.1M_\odot$ for  15\% and 30\% mass losses}
\label{fig:mass_spectra}
\end{figure}

 In Fig 2 and 3, we show the mass spectrum $P(M)$ and the mass loss spectrum as a function of initial mass initial mass. The $P(M)$ spectrums are compared, assuming a starting masses of $m^* = 0.05 M_\odot$ and $m^* = 0.1 M_\odot$. and where the initial probability is calculated assuming  $P(M) = (m^*/M')/log(M_{max}/m^*)$ and $M_{max} = 10M_\odot$. Fig. 2 and 3 suggests that the largest change between the initial and final PBH mass spectrum is in the $10^{-2}$ – $10^{0} M_\odot$. Indeed, our mass-loss mechanism leads to a predicted low-mass cutoff of the PBH mass spectrum at $\lesssim10^{-2}M_{\odot}$. Therefore any observational constraints on sub-stellar mass dark compact objects may provide information relating to the origin of dark matter under our mass-loss models. 

As we noted above the initial redshifts for our calculation were chosen to allow the 
abundance of PBHs with masses $>1000 M_\odot$ to match the observed abundance of seeds for galaxies. The PBH masses were then chosen to be consistent with the horizon mass

\begin{equation}
        M_{H} = 4\pi\rho R_{H}^{3}
\end{equation}

\noindent where $R_{H}$ is the distance to the horizon (= 3t in a radiation dominated universe). This mass is relevant because in the early universe BHs tend to form at the horizon \citep{Chapline1975}. The initial PBH mass can then be estimated using the extrapolated density of dark matter and the frequency for quanta where the event horizon becomes opaque to thermal radiation \citep{Chapline2001}:

\begin{equation}
                \nu = 0.3\text{GeV}\left(\frac{M_{o}}{M_{BH}}\right)^{1/2}
\end{equation}

\noindent In other words when the extrapolated CMB temperature reaches $\approx 1$ GeV, then a PBH can radiate all of its mass attributable to its internal entropy \citep{Chapline2001}. It is interesting that the OGLE microlensing data does include several events with apparent masses below 1 solar mass \citep{Perkins2024}, although it may turn out that these events represent brown dwarfs rather than PHs. It is interesting that analysis of the OGLE microlensing data does include several events with masses apparently below $1M_\odot$ \citep{Perkins2024}. At the present time it is unclear whether these events represent PBHs or "brown dwarfs".
\section{Observational Prospects}

Photometric microlensing surveys \citep[e.g.,][]{Udalski2015} currently provide the tightest constraints on compact dark object dark matter in the sub-stellar mass ranges. Observations of short-timescale photometric microlensing events towards the Galactic Bulge \citep{Niikura2019OGLE} and M31 \citep{Niikura2019} place constraints on the fraction of dark matter in PBHs at $\lesssim10\%$ and $\lesssim1\%$ for the mass ranges [$10^{-6}$-$10^{-3}$]$M_{\odot}$ and [$10^{-11}$-$10^{-6}$]$M_{\odot}$, respectively. However, the constraints towards the Bulge are sensitive to whether short-scale timescale events are assumed to be caused by PBHs or a population of free floating planets with similar masses \citep{Niikura2019OGLE}. Constraints from the microlensing surveys towards the Magellanic Clouds \citep[e.g.,][]{Tisserand2007,Wyrzykowski2011} constrain the remaining substellar mass-range  [$10^{-3}$-$10^{0}$]$M_{\odot}$ ruling out the fraction of dark matter in PBHs at $\lesssim 5\%$. Although the current microlensing constraints rule out a large fraction of dark matter being in form of PBHs \citep[e.g.,][]{Bird2023}, they are not sufficient to definitively probe a sub-stellar PBH mass spectrum cutoff. \\

Prospects for probing the sub-stellar PBH mass spectrum will improve with upcoming microlensing surveys. In addition to increased photometric sensitivity, for example the Vera C. Rubin Observatory \citep{Ivezic2019} could provide constrains on the fraction of dark matter in compact objects down to 
$\lesssim0.1\%$ for masses in the range [$10^{-3}$-$10^{0}$]$M_{\odot}$ with photometric microlensing \citep{Drlica-Wagner2019}, however this is strongly dependant on Galactic Bulge observing cadences chosen \citep{Abrams2023}. Future surveys will also be able to constraint more powerful microlensing observables. Simultaneous photometric monitoring of events from spatially separated observatories such as the Roman Space Telescope \citep[RST;][]{Spergel2015} at the second Lagrange point and ground-based surveys \citep[e.g., PRIME;][]{Kondo2023} will allow microlensing parallax to be measured for the free-floating planet mass ranges \citep{Refsdal1966,Gould2021} which could provide the means to disentangle PBHs from a free-floating planet population \citep{Pruett2022, Lam2020, Perkins2024}. \\

Finally, the advent of sub-mas astrometric capable observatories (e.g., RST and Gaia; \citealt{Prusti2016}), enables the astrometric signatures of microlensing events \citep{Walker1995, Hog1995, Miyamoto1995} to be detected which can provide direct lens-mass measurements \citep[e.g.,][]{Sahu2022, Lam2022,McGill2023} leading to direct constraints across the dark matter mass spectrum \citep{VanTilburg2018, Verma2023, Mondino2023, Fardeen2023}.

\begin{acknowledgments}
We would like to thank Peter McGill for pointing out that due to the emergence of new observing platforms it will be possible to confirm the exitence of a lower bound on PBH masses in the near future and We would like to thank Scott Perkins for comments regarding the interpretation of the OGLE microlensing data as well as the compatibility of our results with the LIGO observations. One of us (GC) would like to thank the Lawrence Livermore National Laboratory for financial support under Contract DE.AC52-07NA27344. 
\end{acknowledgments}

\bibliography{ref}{}
\bibliographystyle{aasjournal}

\end{document}